\documentclass[letterpaper,10pt,journal]{IEEEtran}  
\IEEEoverridecommandlockouts                              
\usepackage{graphicx}
\usepackage{color}
\usepackage{epsfig}
\usepackage{amsmath}
\usepackage{amssymb}
\usepackage{verbatim}
\usepackage{lipsum}
\usepackage{hyperref}
\usepackage{cleveref}
\usepackage{relsize}
\usepackage{epstopdf}
\usepackage{enumerate}
\usepackage{multicol,afterpage,wrapfig}
\usepackage{relsize}
\usepackage{amsthm}
\usepackage{mathtools}
\usepackage{cite}
\usepackage{subcaption}

\usepackage{balance}

\usepackage{csquotes}
\usepackage[table]{xcolor}
\usepackage{tabularx}
\usepackage{multirow}
\definecolor{lightgray}{gray}{0.9}
\allowdisplaybreaks 
\def\matern{{Mat{\'e}rn\;}}
\newcommand{\myeq}[1]{\mathrel{\overset{\makebox[0.07pt]{\mbox{(#1)}}}{=}}}

\def\nb0{{\mathbf{0}}}
\def\nb1{{\mathbf{1}}}

\def\ncalB{{\mathcal{B}}}

\def\ncalI{{\mathcal{I}}}

\def\ncalK{{\mathcal{K}}}

\def\ncalM{{\mathcal{M}}}
\def\ncalN{{\mathcal{N}}}

\def\nbbE{{\mathbb{E}}}

\def\nbbN{{\mathbb{N}}}

\def\nbbP{{\mathbb{P}}}

\def\nbbR{{\mathbb{R}}}

\def\nbbZ{{\mathbb{Z}}}

\def\sinc{{\rm sinc}}

\newtheorem{lemma}{Lemma}

\newtheorem{ndef}{Definition}

\newtheorem{theorem}{Theorem}

\def\figref#1{Fig.\,\ref{#1}}%
\def\E{\mathbb{E}}

\def\pc{\mathtt{P_c}}

\def\sir{\mathtt{SIR}}

\usepackage{stackengine}

\graphicspath{{./Fig/}}
\title{Poisson Cluster Process: Bridging the Gap Between PPP and 3GPP HetNet Models
}

\author{Chiranjib Saha, Mehrnaz Afshang, and Harpreet S. Dhillon
\thanks{The authors are with Wireless@VT, Department of ECE, Virginia Tech, Blacksburg, VA, USA. Email: \{csaha,  mehrnaz,  hdhillon\}@vt.edu. The support of the US National Science Foundation (Grants CCF-1464293 and CNS-1617896) is gratefully acknowledged. 
} }

\let\emptyset\varnothing

\begin{document}

\maketitle
\thispagestyle{empty}
\pagestyle{empty}
\vspace{-2em}
\begin{abstract}
The growing complexity of heterogeneous cellular networks (HetNets) has necessitated the need to consider variety of user and base station (BS) configurations for realistic performance evaluation and system design. This is directly reflected in the HetNet simulation models considered by standardization bodies, such as the third generation partnership project (3GPP). Complementary to these simulation models, stochastic geometry-based approach modeling the user and BS locations as independent and homogeneous Poisson point processes (PPPs) has gained prominence in the past few years. Despite its success in revealing useful insights, this PPP-based model is not rich enough to capture all the spatial configurations that appear in real-world HetNet deployments (on which 3GPP simulation models are based). In this paper, we bridge the gap between the 3GPP simulation models and the popular PPP-based analytical model by developing a new unified HetNet model in which a fraction of users and some BS tiers are modeled as Poisson cluster processes (PCPs). This model captures both non-uniformity and coupling in the BS and user locations. For this setup, we derive exact expression for downlink coverage probability under maximum signal-to-interference ratio ($\sir$) cell association model. As intermediate results, we define and evaluate {\em sum-product functionals} for PPP and PCP. Special instances of the proposed model are shown to closely resemble different configurations considered in 3GPP HetNet models. Our results  concretely demonstrate that the performance trends are highly sensitive to the assumptions made on the user and SBS configurations. 
\end{abstract}  
\vspace{-0.9em}
\section{Introduction}
In order to handle exponential growth of mobile data traffic, macrocellular networks of the yesteryears have gradually evolved into more denser heterogeneous cellular networks in which several types of low power small cells coexist with macrocells. While macro BSs (MBSs) were deployed fairly uniformly to provide a ubiquitous coverage blanket, the small cell BSs (SBSs) are deployed somewhat organically by operators to complement coverage and capacity of the cellular networks at user hotspots, or by subscribers or operators to patch coverage dead-zones. This naturally couples the locations of the SBSs with the users, as a result of which we now need to consider plethora of deployment scenarios in the system design phase as opposed to only a few in the macro-only networks of the past. As discussed in detail in Section~\ref{sec::hetnet_model}, this is also reflected in the 3GPP simulation models which now have to consider several different configurations of user and SBS locations in the system-level simulations. For instance, in addition to the uniform user locations, 3GPP simulation models also consider clustered configurations in which the user and SBS locations are coupled. 

The inherent irregularity in the SBS locations also motivated a complementary analytical approach based on modeling the user and BS locations by point processes. This allows the use of powerful tools from stochastic geometry to facilitate tractable analysis of key network performance metrics, such as coverage and rate. However, as discussed in Section~\ref{sec::hetnet_model} in detail, the existing analytical model (first proposed in~\cite{5743604,dhillon2012modeling}) models the user and BS locations using independent homogeneous PPPs in order to maintain tractability. Although this PPP-based HetNet model has yielded significant insights into the network behavior, it is not rich enough to emulate all user and SBS configurations that appear in the real-world deployments (on which 3GPP simulation models are based). 

In our recent works, we have explored some specific instances of the 3GPP simulation models and argued that PCP can potentially bridge the gap between these instances of the 3GPP models and the PPP-based baseline analytical HetNet model by incorporating the clustering effect of points which naturally appears in the locations of users (due to hotspot formation) and SBSs (due to deployment at the user hotspots) \cite{SahaAfshDh2016,AfshDhiClusterHetNet2016,DengZhouHaenggi2015,HetHetNets2015,AfshSahDhi2016Contact}. We will discuss the instances studied in these works in the next Section. In this paper, we first provide a brief overview of the spatial configurations of the BSs and users that are considered in the 3GPP simulation models of HetNets. We then develop a new unified PCP-based HetNet model, which accurately captures all these configurations as its special instances. Further details are provided next.


 {\em Contributions and Outcomes.} We propose a general and flexible  analytical framework for $K$-tier HetNet where (i) the users are either distributed as homogeneous PPP 
 or a PCP, (ii) the BS tiers can be either PPP or PCP coupled to the user point process. All SBS and user configurations considered by 3GPP can be interpreted as special instances of this general model. For this model, we derive the  downlink coverage under the max-$\sir$ cell association. As a part of this analysis, we first introduce a family of sum-product functionals for PPP and PCP. Explicit expressions for these functions are then derived (for both PPP and PCP). We then show that the coverage probability for this setup can be expressed as a summation of these sum-product functionals. Finally, we specialize the coverage probability expressions for a family of PCP, known as Neyman Scott processes. In numerical result section, we further specialize the general multi-tier setup to different configurations considered in 3GPP simulation models. Our results concretely demonstrate that the performance trends are highly sensitive to the assumptions made on the user and SBS configurations, which further highlights the importance of the proposed all-inclusive HetNet model.
   \begin{figure}
   \centering
  \begin{subfigure}{.30\textwidth}
  \centering
  \includegraphics[width=\linewidth]{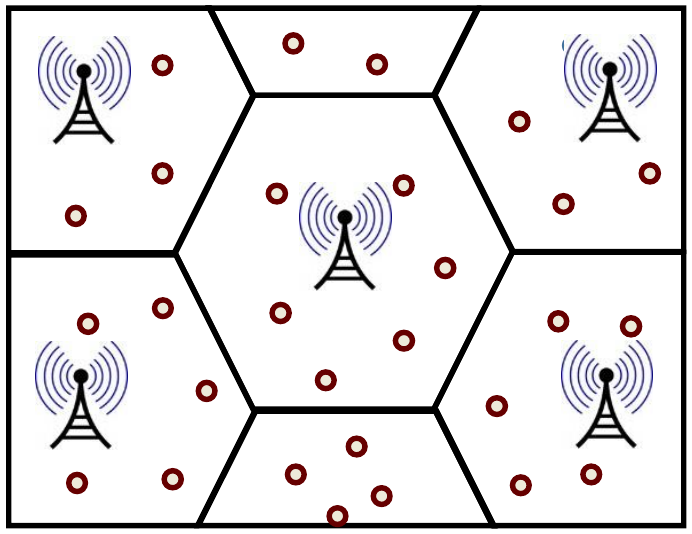}
  \caption{}
  \label{fig:user_uniform_3gpp}
\end{subfigure}%
\hspace{0.2cm}
\begin{subfigure}{.30\textwidth}
  \centering
  \includegraphics[width=\linewidth]{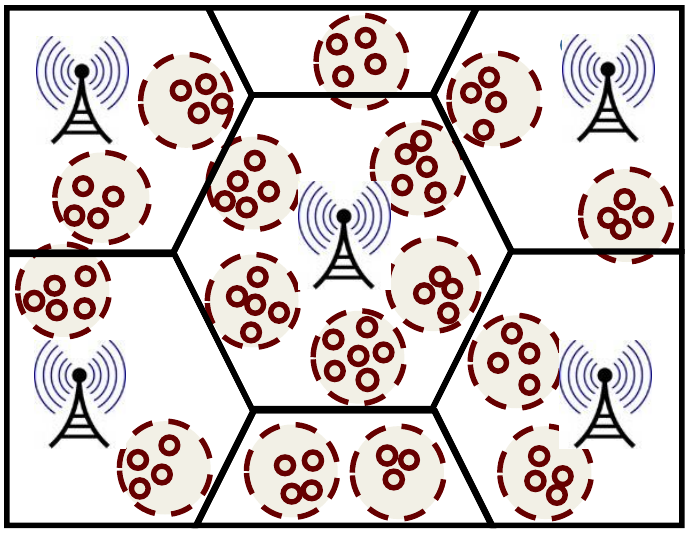}
  \caption{}
  \label{fig:user_clustered_3gpp}
\end{subfigure}
\caption{\small User configurations in 3GPP HetNet model: (a) ``uniform'' users within a  macro cell, and (b) ``clustered'' users within a  macro cell.}\label{fig::user_config::3gpp}
\end{figure}
  \begin{table*}
\caption{\small 3GPP Model Configuration for HetNets (source \cite{3GPPheterogeneous2011}).}\label{table::3gpp_configuration}
\begin{center}
\rowcolors{1}{}{lightgray}
\begin{tabular}{c c c c l}
  \hline
 Configuration & \vtop{\hbox{\strut  User Density}\hbox{\strut Across macro cells}}
 & \vtop{\hbox{\strut User Distribution} \hbox{\strut within a macro cell}} & \vtop{\hbox{\strut SBS distribution}\hbox{\strut within a macro cell}} &Comments\\
 1&Uniform&Uniform&Uncorrelated&Capacity enhancement\\
 2&Non-uniform&Uniform & Uncorrelated & Sensitivity to non-uniform user density across macro cells\\
 3 &Non-uniform &Uniform &Correlated & Cell edge enhancement\\
 4 &Non-uniform &Clustered &Correlated &Hotspot capacity enhancement\\
  \hline
\end{tabular}
\end{center}
\end{table*}
\section{HetNet Models}\label{sec::hetnet_model}
In this section, we summarize different classes of spatial models for HetNets that are used by industry (specifically 3GPP) and academia. We begin by summarizing the models used for system-level simulations by 3GPP. For modeling macrocells, 3GPP simulation scenarios rely on either a single macro cell setup or grid based models, where finite number of MBSs are placed as regularly spaced points on a plane. On the contrary, as discussed next, several different configurations corresponding to variety of real-life deployment scenarios are considered for modeling the locations of users and SBSs (usually pico and femto cells) \cite[Section~A.2.1.1.2]{access2010further}. In all the discussions corresponding to 3GPP scenarios, we will intentionally use {\em keywords} reserved for referring to these configurations in the 3GPP documents. 

{\em Users.} The spatial distribution of users within a macro cell is either \enquote{uniform}\footnote{3GPP documents have an alternative interpretation of ``uniform'' users: it means number of users per macro cell is the same. Otherwise, the users are ``non-uniform'' meaning that different macro cells have different number of users. This differentiation does not appear in our discussion.} (i.e. homogeneous) or \enquote{clustered} forming hotspots (see \figref{fig::user_config::3gpp} for an illustration). Thus, a fraction of  users in a macro cell  are uniformly distributed  inside the cell, and the rest are  clustered around the SBS locations, more specifically  are distributed uniformly at random within a circular region of constant radius centered around each SBS.  

\begin{figure}
\centering
\includegraphics[width =.3 \textwidth]{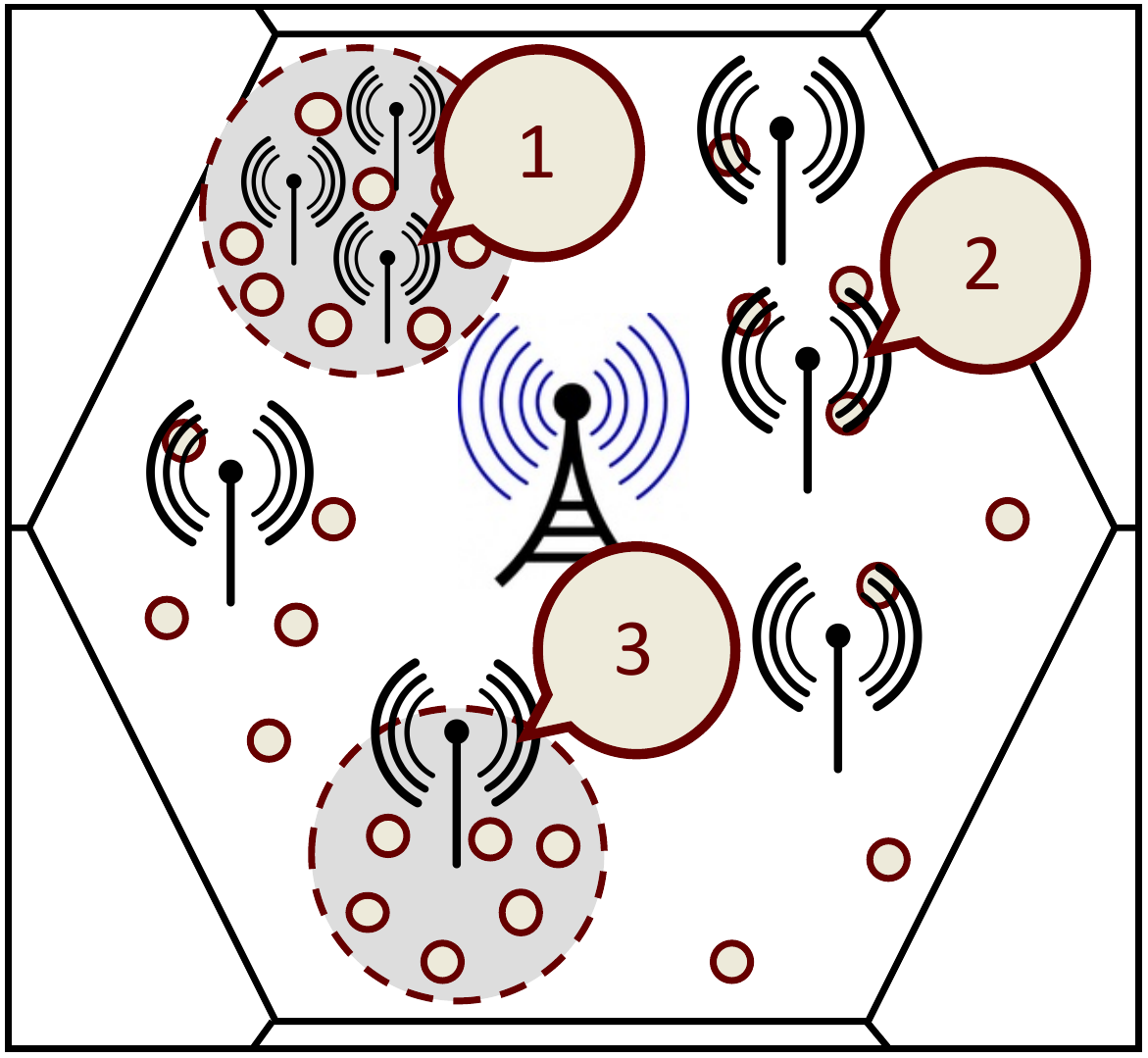}
\caption{\small Spatial distribution of SBSs in 3GPP HetNet model: (1) SBSs deployed at higher density at certain areas (indoor models), (2) SBSs deployed randomly  or under some site  planning,  and (3) a single SBS deployed at the center of a user hotspot.
 }\label{fig::sbs::3gpp}
\end{figure}

{\em Base stations.} The MBSs are placed deterministically in a grid. The SBS locations inside a macro cell are either  \enquote{uncorrelated} (meaning that they  are  distributed uniformly at random inside a macro cell) or \enquote{correlated}. This {\em correlation} is induced by different site planning optimization strategies, such as (i)  more SBSs are deployed at user hotspots for capacity enhancement, and (ii) some locations at cell edge are selected for SBS placement to maximize the cell edge coverage.  See \figref{fig::sbs::3gpp} for an illustrative example.     

These configurations of user and BS placements provide a rich set of combinations to study variety of real-word deployment scenarios for HetNets. These different combinations are summarized in Table~\ref{table::3gpp_configuration}, which appears in \cite{3GPPheterogeneous2011}. 


\begin{figure}
\begin{subfigure}{.24\textwidth}
\includegraphics[width=\columnwidth]{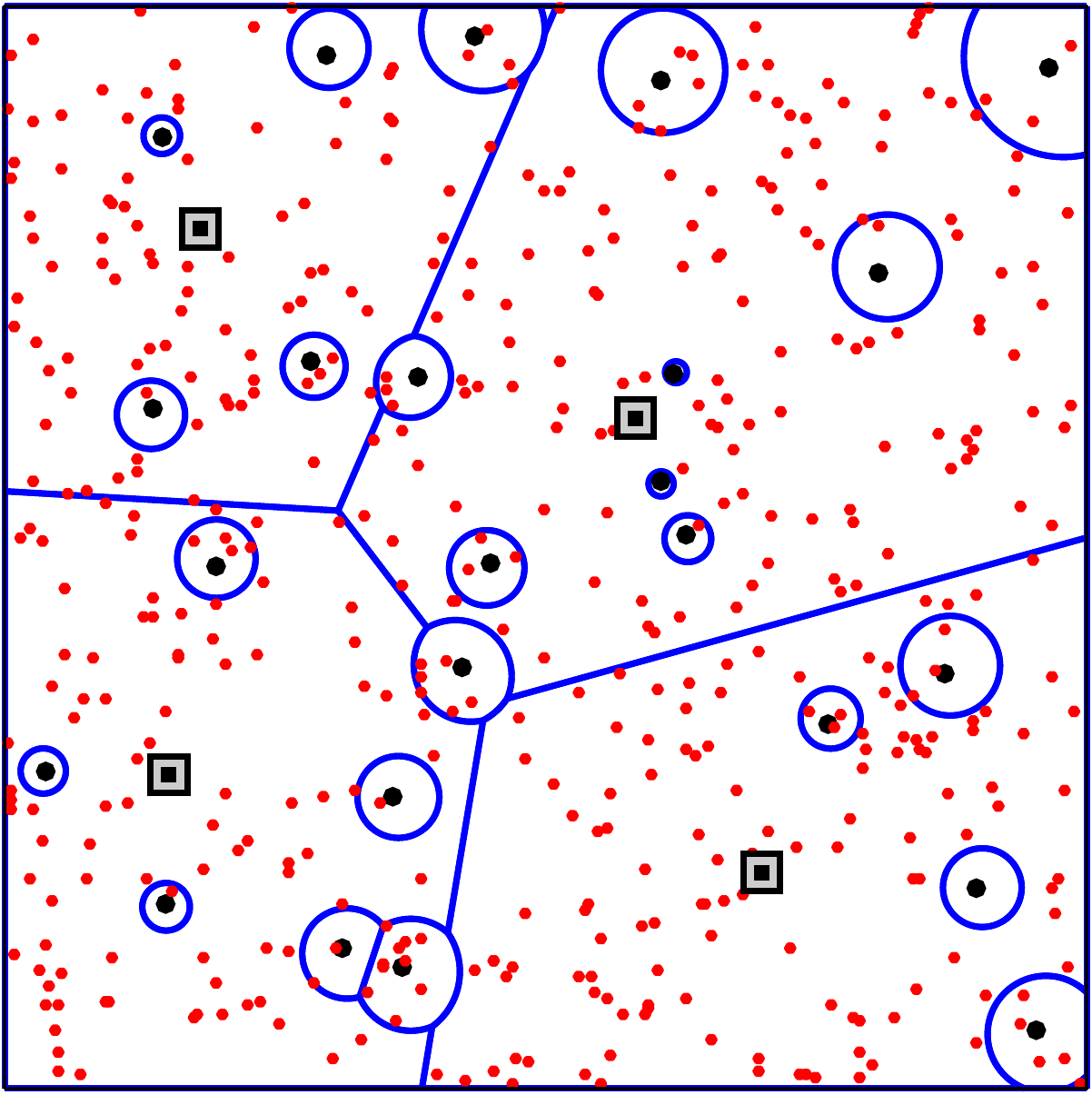}
  \caption{Model 1: SBS PPP, user PPP}
  \label{fig:sfig1}
\end{subfigure}
\begin{subfigure}{.24\textwidth}
\includegraphics[width=\columnwidth]{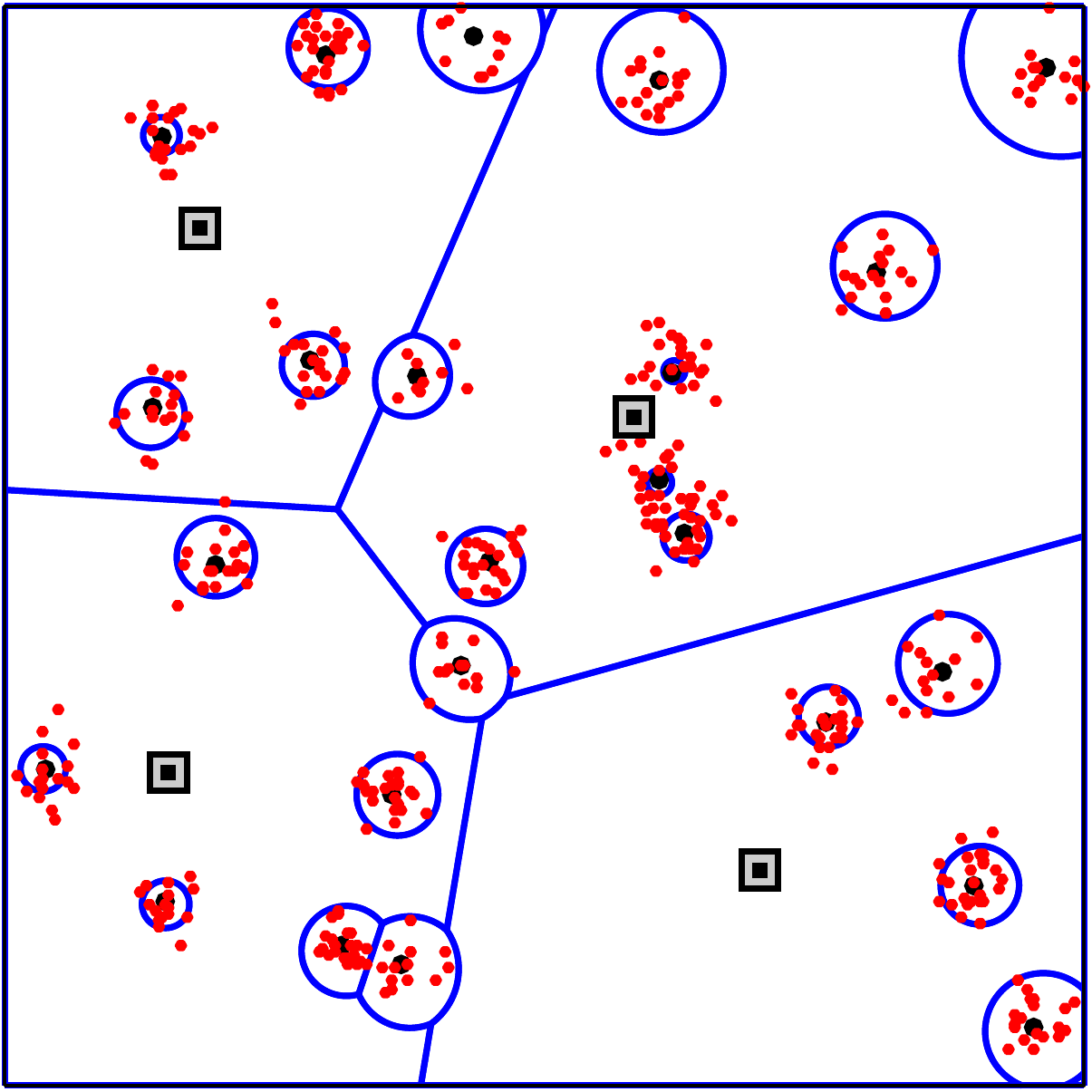}
  \caption{Model 2: SBS PPP, user PCP}
  \label{fig:sfig2}
\end{subfigure}
\begin{subfigure}{.24\textwidth}
\includegraphics[width=\columnwidth]{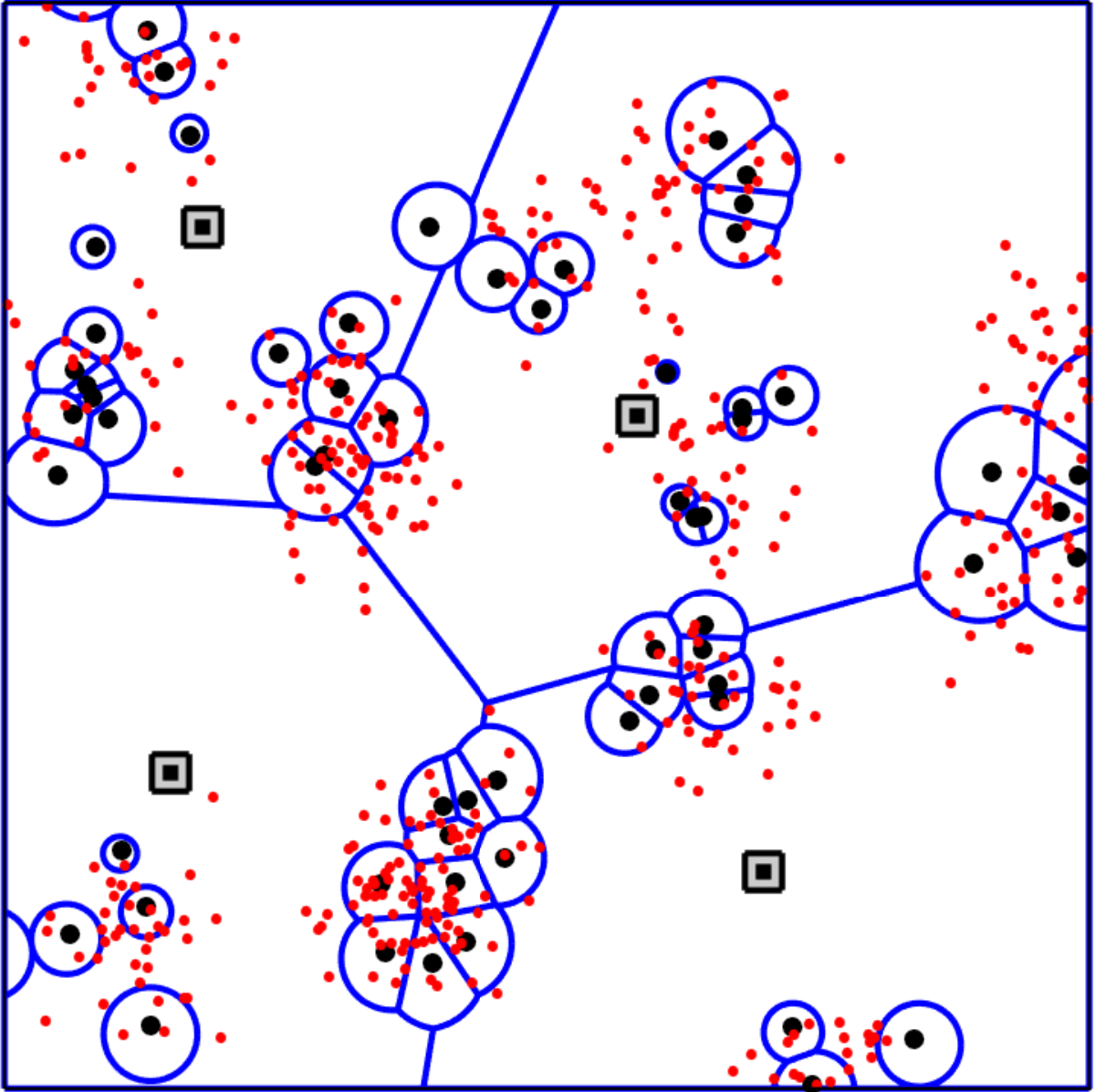}
  \caption{Model 3: SBS PCP, user PCP}
  \label{fig:sfig3}
\end{subfigure}
\begin{subfigure}{.24\textwidth}
\includegraphics[width=\columnwidth]{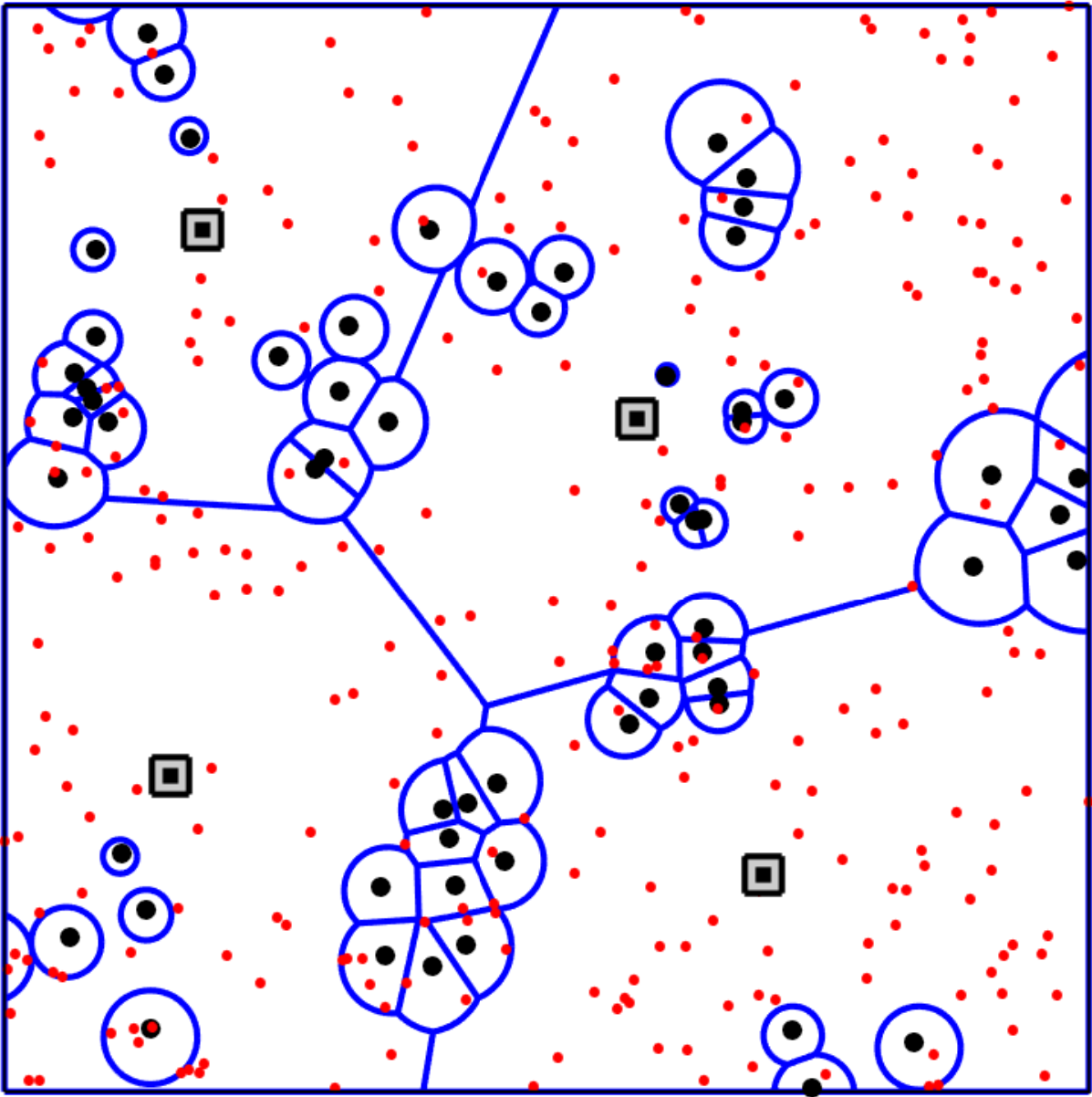}
  \caption{Model 4: SBS PCP, user PPP}
  \label{fig:sfig4}
\end{subfigure}
\caption{\small Illustration of the four generative models with the combination of PPP and PCP. The black square, black dot and red  dot refer to the macro BS, SBS and users respectively.}
\label{fig::hetnet::model}
\end{figure}

In parallel to these efforts by 3GPP, {\em analytical} HetNet models with foundations in stochastic geometry have gained prominence in the last few years~\cite{elsawy2013stochastic,andrews2016primer,mukherjee2014analytical}. The main idea here is to endow the locations of the BSs and users with distributions and then use tools from stochastic geometry to derive easy-to-compute expressions for key performance metrics, such as coverage and rate\footnote{A careful reader will note that 3GPP models also endow the locations of users and small cells with distributions, which technically makes them stochastic models as well.}. In order to maintain tractability, the locations of the users and different types of BSs are usually modeled by independent homogeneous PPPs~\cite{elsawy2013stochastic,andrews2016primer,mukherjee2014analytical}. We will henceforth refer to homogeneous PPP as a PPP unless stated otherwise. This model is usually referred to as a $K$-tier HetNet model and was first introduced in~\cite{5743604,dhillon2012modeling}. Despite the simplicity of a PPP, this model is known to be a reasonable choice for the locations of MBSs~\cite{AndrewsTractable}, users that are distributed uniformly over the plane, as well as the SBSs that are located uniformly at random over a macrocell. Therefore, roughly speaking, this model is capable of accurately modeling configuration 1 from Table~\ref{table::3gpp_configuration}. However, this PPP-based model is not rich enough to capture non-uniformity and coupling across the locations of the users and SBSs (such as in configurations 2-4 in Table~\ref{table::3gpp_configuration})~\cite{NonUniformDhillon,SahaAfshDh2016,AfshDhiClusterHetNet2016,DengZhouHaenggi2015}. In order to capture that accurately, we need to model these locations using point processes that exhibit inter-point attraction. A simple way of achieving that, which is also quite consistent with the 3GPP configurations listed in Table~\ref{table::3gpp_configuration}, is to use PCPs~\cite{ganti2009interference,AfshDhi2015MehrnazD2D1}. By combining PCP with a PPP, we can create generative models that are rich enough to model different HetNet configurations of Table~\ref{table::3gpp_configuration}. We discuss these generative models next. 
 
\begin{itemize}
\item \textit{Model 1: SBS PPP, user PPP.} This is the PPP-based $K$-tier baseline model most commonly used in HetNet literature and is in direct agreement with the 3GPP models with \textit{uniform} user and \textit{uncorrelated} SBS distribution; see \cite{5743604,dhillon2012modeling,mukherjee2012distribution,jo2012heterogeneous,MadhusudhananRestrepoBrown2016} for a small subset.  
\item \textit{Model 2: SBS PPP, user PCP.} Proposed in our recent work~\cite{SahaAfshDh2016}, this model can accurately characterize \textit{clustered} users and \textit{uncorrelated} SBSs.  In particular, we model the clustered user and SBS locations jointly  by defining PCP of users around PPP distributed SBSs. This captures the coupling between user and SBS locations.
\item \textit{Model 3: SBS PCP, user PCP.}  As discussed already, the SBS locations may also be correlated  and form spatial clusters according to the user hotspots for capacity-centric deployment. For such scenario, two PCPs with the same parent PPP but independently and identically distributed (i.i.d.) offspring point processes are  used to model the users and SBS locations. Coupling is modled by having the same parent PPP for both the PCPs. We proposed and analyzed this model for HetNets in~\cite{AfshDhiClusterHetNet2016}.
 \item \textit{Model 4: SBS PCP, user PPP.} This scenario can occur in conjunction with the previous one since some of the users may not be a part of the user clusters but are still served by the clustered SBSs. PPP is  a good choice for modeling user locations in this case~\cite{HetPCPGhrayeb2015, MultiChannel2013}. 
\end{itemize}   
These generative models are illustrated in \figref{fig::hetnet::model}. In the next section, we will develop a unified model (of which these generative models will be special instances), which will significantly enhance the PPP-based $K$-tier model of \cite{5743604,dhillon2012modeling}. 

\section{System Model}\label{sec::system_model}
We assume a $K$-tier HetNet consisting of $K$ different types of BSs distributed as PPP or PCP. In particular, we denote the point process of the $k^{th}$ BS tier as $\Phi_k$,  where $\Phi_k$ is either a PPP with density $\lambda_k$ ($\forall  k\in\ncalK_1$) or a PCP ($\forall k\in\ncalK_2$), where $\ncalK_1$ and  $\ncalK_2$ are the index sets  of the BS  tiers being modeled as PPP and PCP,  respectively, with  $|\ncalK_1\cup\ncalK_2| = K$. 

A PCP $\Phi_k$ can be uniquely defined as:
\begin{equation}
{\Phi_k = \bigcup_{{\bf z}\in \Phi_{{\rm p}_k}} {\bf z} + {\cal B}^{\bf z}_k,}
\end{equation} 
where $\Phi_{{\rm p}_k}$ is the parent PPP of density $\lambda_{{\rm p}_k}$ and $\ncalB_k^{\bf z}$ denotes the offspring point process where each point at ${\bf s}\in\ncalB_k^{\bf z}$ is  i.i.d. around the cluster center ${\bf z}\in\Phi_{{\rm p}_k}$ with density $f_k({\bf s})$. If $N_k $ denotes the number of points in $\ncalB_k^{\bf z}$, then $N_k\sim p_{k(n)}$ ($n\in\nbbN$). While we put no restriction on $p_{k(n)}$ for the coverage probability analysis, we later specialize our results for  Neyman Scott processes where  $N_k\sim\mathtt{Poisson}(\bar{m}_k)$. An illustration of the realization of this process is provided in \figref{fig::matern}.

We assume that each BS of $\Phi_k$ transmits at constant power $P_k$. 
Define $\Phi_{\rm u}$ as the user point process.  We perform our analysis for a {\em typical} user which corresponds to a  point selected uniformly at random from $\Phi_{\rm u}$. Without loss of generality, the typical user is located at origin. Contrary to the common practice in the literature,   $\Phi_{\rm u}$ is not necessarily a PPP independent of the BS locations, rather this scenario will appear as a special case in our analysis. In particular, we consider three different configurations for users: 
\begin{figure}
\centering
\includegraphics[width =0.28 \textwidth]{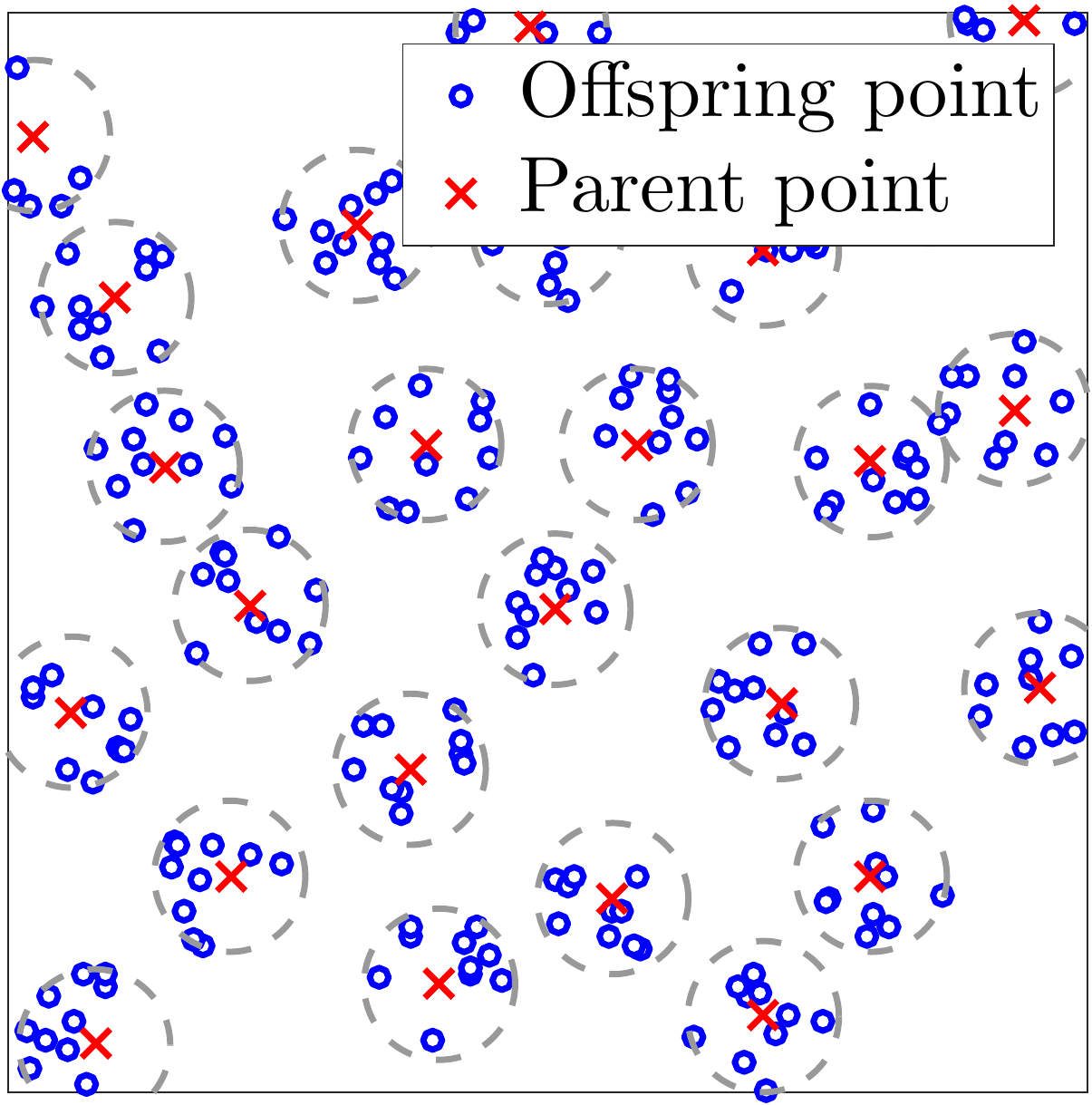}
\caption{\small A realization of a \matern cluster process: a special case of Neyman Scott process where the offspring points are distributed uniformly inside a disc around the cluster center.}\label{fig::matern}
\end{figure}
\begin{itemize}
\item Case 1 ({\em uniform users}): $\Phi_{\rm u}$ is  a  PPP. This corresponds to Models $1$ and $4$ from the previous Section.

\item Case 2  ({\em clustered users}): $\Phi_{\rm u}$ is a PCP with parent PPP $\Phi_i$ ($i\in \ncalK_1$), which corresponds to Model $2$ (single SBS deployed in a user hotspot).

\item Case 3 ({\em clustered users}): $\Phi_{\rm u}$ is a PCP having same parent PPP as that of $\Phi_i$ ($i\in\ncalK_2$), which corresponds to Model $3$ (multiple SBSs deployed at a user hotspot).
\end{itemize}

Since the locations of the users and BSs are coupled in cases 2 and 3, when we select a typical user, we also implicitly select the cluster to which it  belongs. It will be useful to separate out the BSs located in this cluster from the rest of the point process, which we do next. For case 2, let ${\bf z}_0\in\Phi_i\ (i\in\ncalK_1)$ is the location of the BS at the  cluster center of the typical user. For  case 3, let us define the {\em representative} BS cluster $\ncalB^{{\bf z}_0}_{i}\subset \Phi_i\  (i\in\ncalK_2)$ having the  cluster center ${\bf z}_0$ which is also the cluster center of the typical user located at origin. 
Having defined all three possible configurations/cases of $\Phi_{\rm u}$, we define a set
\begin{align}
\Phi_0 =\begin{cases} \emptyset;&\text{case 1,}\\
\{{\bf z}_0\};&\text{case 2,}\\
\ncalB^{{\bf z}_0}_i;&{\text{case 3.}}
\end{cases}
\end{align}
This set can be interpreted as the locations of the BSs that lie in the same cluster as the typical user. By Slivnyak's theorem, we can remove $\Phi_0$ from $\Phi_i$ without changing the distribution of $\Phi_i$. Therefore, for case 2, we remove singleton $\{{\bf z}_0\}$ from $\Phi_i (i\in\ncalK_1)$ whereas in case 2, we remove finite process $\ncalB^{{\bf z}_0}_i$, which is a representative cluster of BSs with properties ($ f_i(\cdot), N_i$) being inherited from $\Phi_i$ ($i\in\ncalK_2$). Note that since $\Phi_0$ is constructed from $\Phi_i$ ($i\in\ncalK_1\cup\ncalK_2$), the transmit power of the BS(s) is $P_0\equiv P_i$. 
Hence, 
the BS point process is a superposition of independent point processes defined as:
\begin{equation}
{\Phi =   \bigcup\limits_{k_1\in\ncalK_1}\Phi_{k_1}\bigcup\limits_{k_2\in\ncalK_2}\Phi_{k_2}\bigcup\Phi_0,}
\end{equation}
and the corresponding index set is enriched  as: $\ncalK= \ncalK_1\cup\ncalK_2\cup\{0\}$. 
For the ease of exposition, the thermal noise is assumed to be  negligible compared to the  interference power.
Assuming the serving BS is located at ${\bf x}\in\Phi_k$, $\sir({\bf x})$ is defined as:
\begin{align}\label{eq::sir::definition}
\sir({\bf x}) = \frac{P_kh_{\bf x}\|{\bf x}\|^{-\alpha}}{{\ncalI(\Phi_k\setminus\{{\bf x}\})+\sum\limits_{j\in\ncalK\setminus\{k\}}\ncalI(\Phi_j)}},
\end{align} 
where $\ncalI(\Phi_i)  = \sum_{y\in\Phi_i}P_i h_{\bf y} \|{\bf y}\|^{-\alpha}$ is the aggregate interference from $\Phi_i$ ($i\in\ncalK$). 
For the channel model, we assume signal from a BS at ${\bf y}\in\nbbR^2$ undergoes independent Rayleigh fading, more precisely  $\{h_{\bf y}\}$ is an i.i.d. sequence of random variables, with $h_{\bf y}\sim \exp(1)$ and $\alpha > 2$ is the path-loss exponent. Assuming $\beta_k$ is the $\sir$-threshold defined for $\Phi_k$ for successful connection and the user connects to the BS  that provides maximum $\sir$,  coverage probability is defined as: 
\begin{align}\label{eq::coverage_definition}
\pc &= \nbbP\bigg[\bigcup\limits_{k\in\ncalK}\bigcup\limits_{{\bf x}\in\Phi_k}\{ \sir({\bf x})>\beta_k \}\bigg].
\end{align}

Note that $\beta_0 \equiv \beta_i$ for cases 2 and 3 defined above. 
In the next Section, we derive  the main result for the coverage probability of the typical user under the assumption that all BSs are operating in open access.

\section{Coverage Probability Analysis}

Before going into the coverage probability analysis, we derive key intermediate results. These results, such as sum product functionals of PCPs and finite processes, are useful on their own right.

\subsection{Sum-Product Functionals}\label{sec::cov::prob::sum::pr}
 We first  define sum-product functional over point processes,  which will be useful for the derivation of the coverage probability expression under max-$\sir$ connectivity. 
\begin{ndef}[Sum-product functional] Sum-product functional of a point process $\Psi$ can be defined as:
\begin{align}\label{eq::sum::product::defn}
\nbbE\left[\sum\limits_{{\bf x}\in{\Psi}}g({\bf x})\prod\limits_{{\bf y}\in\Psi\setminus\{{\bf x}\}}v({\bf x},{\bf y})\right],
\end{align}
where $g({\bf x}):\nbbR^2\mapsto[0,1]$ and $v({\bf x},{\bf y}):[\nbbR^2\times\nbbR^2]\mapsto[0,1]$ are measurable. 
\end{ndef}
In the following Lemma, we provide the expression for sum-product functional when $\Psi$ is a PPP. 
\begin{lemma}\label{lemm::sumproduct::ppp}The sum-product functional of $\Psi$ when $\Psi$ is a PPP (i.e., $\Psi \equiv \Phi_k (k\in\ncalK_1)$) can be expressed as follows: 
\begin{align}\label{eq::sumproduct::ppp}
\nbbE\left[\sum\limits_{{\bf x}\in{\Psi}}g({\bf x})\prod\limits_{{\bf y}\in\Psi\setminus\{{\bf x}\}}v({\bf x},{\bf y})\right]&=\lambda_k\int\limits_{\nbbR^2} g({\bf x}){G}(v({\bf x},{\bf y})){\rm d}{\bf x},
\end{align}
where ${G}(\cdot)$ denotes the probability generating functional (PGFL) of $\Psi$. 
\end{lemma}
\begin{IEEEproof}
We can directly apply Campbell Mecke Theorem \cite{chiu2013stochastic} to evaluate \eqref{eq::sum::product::defn} as:
\begin{align*}
&\nbbE\bigg[\sum\limits_{{\bf x}\in\Psi}g({\bf x})\prod\limits_{{\bf y}\in\Psi\setminus\{{\bf x}\}}v({\bf x},{\bf y})\bigg]
\\& =  \int\limits_{\nbbR^2}g({\bf x})\nbbE_{\bf x}^! \prod\limits_{{\bf y}\in\Psi}v({\bf x},{\bf y})\Lambda({\rm d}{\bf x}) = \int\limits_{\nbbR^2} g({\bf x})\widetilde{G}(v({\bf x},{\bf y}))\Lambda({\rm d}{\bf x}),
\end{align*}
where $\Lambda(\cdot)$ is  the intensity measure of $\Psi$ and  $\widetilde{G}(\cdot)$ denotes the PGFL of $\Psi$ under its reduced Palm distribution. When $\Psi$ is  homogeneous PPP, $\Lambda({\rm d}{\bf x})=\lambda_k\:{\rm d}{\bf x}$ and $\widetilde{G}(v({\bf x},{\bf y})) =G(v({\bf x},{\bf y})) = \nbbE\prod\limits_{{\bf y}\in\Psi} v({\bf x},{\bf y})$, by Slivnyak's theorem~\cite{chiu2013stochastic}.
\end{IEEEproof}
Sum-product functional of $\Psi$ when $\Psi$ is a PCP, i.e., $\Psi \equiv \Phi_k\ (k\in\ncalK_2)$ requires  more careful treatment since selecting a point from ${\bf x}\in\Psi$ implies selecting a tuple $({\bf x},{\bf z})$, where ${\bf z}$ is the cluster center of ${\bf x}$. Alternatively, we can assign a two-dimensional mark ${\bf z}$ to each point ${\bf x}\in\Psi$ such that ${\bf z}$ is the cluster center of ${\bf x}$. Then $({\bf x},{\bf z})$ is a point from the marked point process $\hat{\Psi}\subset\nbbR^2\times\nbbR^2$. 
It should be noted that $\hat{\Psi}$ is simply an alternate representation of $\Psi$, which will be useful in some proofs in this Section. Taking $A,B\subset \nbbR^2$, its intensity measure can be expressed as:
\begin{align*}
&\Lambda(A, B) = \nbbE\bigg[\sum\limits_{({\bf x},{\bf z})\in \hat{\Psi}}{\bf 1}\big({\bf x}\in A, {\bf z}\in B\big)\bigg]\notag\\
&\myeq{a}\nbbE\left[\sum\limits_{{\bf z} \in\Phi_{{\rm p}_k}\cap B}\bar{m}_k\int\limits_{{\bf x}\in A}f_k({\bf x}-{\bf z}){\rm d}{\rm x}\right]\\
&=\bar{m}_k\lambda_{{\rm p}_k}\iint\limits_{{\bf z}\in B,{\bf x}\in A}f_k({\bf x}-{\bf z}){\rm d}{\bf x}{\rm d}{\bf z},
\end{align*}
where in step (a), the expression under summation is the  intensity of ${\bf z}+\ncalB^{{\bf z}}_k$, i.e., the offspring process with cluster center at ${\bf z}$. The last step follows from  the application of Campbell's theorem \cite{chiu2013stochastic}. 
Hence, $\Lambda({\rm d}{\bf x},{\rm d}{\bf z})= \lambda_{{\rm p}_{k}}\bar{m}_kf_k({\bf x}-{\bf z})\:{\rm d}{\bf z}\:{\rm d}{\bf x}$.   
\begin{lemma}\label{lemm::sumproduct::pcp}The sum-product functional of $\Psi$ when $\Psi$ is a PCP (i.e., $\Psi=\Phi_k (k\in\ncalK_2)$) can be expressed as follows: 
\begin{multline}\label{eq::sumproduct::pcp}
\nbbE\left[\sum\limits_{{\bf x}\in{\Psi}}g({\bf x})\prod\limits_{{\bf y}\in\Psi\setminus\{{\bf x}\}}v({\bf x},{\bf y})\right]\\=\iint\limits_{\nbbR^2\times\nbbR^2} g({\bf x})\widetilde{G}(v({\bf x},{\bf y})|{\bf z})\Lambda({\rm d}{\bf x},{\rm d}{\bf z}),
\end{multline}
where
\begin{equation}\label{eq::reduced::palm::pcp}
\widetilde{G}(v({\bf x},{\bf y})|{\bf z}) = {G}(v({\bf x},{\bf y}))\widetilde{G}_c(v({\bf x},{\bf y})|{\bf z})
\end{equation}
 denotes the PGFL of $\Psi$ when a  point ${\bf x}\in\Psi$ with cluster center at ${\bf z}$ is removed from $\Psi$. $G(\cdot)$  is the PGFL of $\Psi$ and $\widetilde{G}_c(\cdot|{\bf z})$ is the PGFL of a cluster of $\Psi$ centered at ${\bf z}$ under its reduced Palm distribution. 
\end{lemma}
\begin{IEEEproof} 
 Starting from \eqref{eq::sum::product::defn} we apply Campbell Mecke theorem on $\hat{\Psi}$ as follows:
\begin{align*}
 &\nbbE\bigg[\sum\limits_{({\bf x},{\bf z})\in\hat{\Psi}}g({\bf x})\prod\limits_{({\bf y},{\bf z}')\in\hat{\Psi}\setminus({\bf x},{\bf z})}v({\bf x},{\bf y})\bigg]\\
 &=
\iint\limits_{\nbbR^2\times \nbbR^2} \nbbE_{({\bf x},{\bf z})}^!\bigg[g({\bf x})\prod\limits_{({\bf y},{\bf z}')\in\hat{\Psi}}v({\bf x},{\bf y})\bigg]\Lambda({\rm d}{\bf x},{\rm d}{\bf z}).
\end{align*}
The Palm expectation in the last step can be simplified as:
\begin{align*}
&\nbbE_{({\bf x},{\bf z})}^!\bigg[g({\bf x})\prod\limits_{({\bf y},{\bf z}')\in\hat{\Psi}}v({\bf x},{\bf y})\bigg]\\&
=g({\bf x})\nbbE\bigg[\prod\limits_{{\bf y}\in\Psi\setminus({\bf z}+\ncalB^{{\bf z}}_k)}v({\bf x},{\bf y})\prod\limits_{{\bf y}\in({\bf z}+\ncalB^{{\bf z}}_k)\setminus\{{\bf x}\}}v({\bf x},{\bf y})\bigg]\\
&=g({\bf x})\nbbE\bigg[\prod\limits_{{\bf y}\in\Psi}v({\bf x},{\bf y})\bigg]\nbbE\bigg[\prod\limits_{{\bf y}\in({\bf z}+\ncalB^{{\bf z}}_k)\setminus\{{\bf x}\}}v({\bf x},{\bf y})\bigg],
\end{align*}
where the last step is obtained by   Slivnyak's theorem \cite{chiu2013stochastic}. Substituting the PGFLs as $\nbbE\prod\limits_{{\bf y}\in\Psi} v({\bf x},{\bf y})=G(v({\bf x},{\bf y}))$, and $ \nbbE\prod\limits_{{\bf y}\in{\bf z}+\ncalB_k^{\bf z}\setminus\{{\bf x}\}}v({\bf x},{\bf y})=\nbbE_{\bf x}^!\prod\limits_{{\bf y}\in{\bf z}+\ncalB_k^{\bf z}}v({\bf x},{\bf y})=\widetilde{G}_c(v({\bf x},{\bf y})|{\bf z}) $, we get the final result.
\end{IEEEproof}

The similar steps for the evaluation of the sum-product functional  can not be followed when $\Psi$ is a finite point process, specifically, $\Psi = {\bf z}+\ncalB^{\bf z}_k$,  the cluster of a randomly  chosen point $ {\bf x} \in \Phi_k$ ($k= 0$) centered at ${\bf z}$.  
\begin{lemma}\label{lemm::sumproduct::finite}The sum-product functional of $\Psi$ when $\Psi$ is the cluster of a randomly  chosen point ${\bf x} \in \Phi_k$ ($k=0$)  with cluster center located at ${\bf z}$
can be expressed as follows: 
\begin{multline}\label{eq::sumproduct::finite}
\nbbE\left[\sum\limits_{{\bf x}\in{\Psi}}g({\bf x})\prod\limits_{{\bf y}\in\Psi\setminus\{{\bf x}\}}v({\bf x},{\bf y})\right]=\sum\limits_{n=1}^{\infty}\int\limits_{\nbbR^2}g({\bf x}) \bigg(\ \int\limits_{\nbbR^2}v({\bf x},{\bf y})\\ \times f_k({\bf y}-{\bf z}){\rm d}{\bf y}\bigg)^{n-1}f_k({\bf x}-{\bf z}){\rm d}{\bf x}\:n^2\frac{p_{k(n)}}{\bar{m}_k}.
\end{multline}
\end{lemma}
\begin{IEEEproof}
Note that $\Psi$ is conditioned to have at least one point (the one located at ${\bf x}$) and the number of points in $\Psi$ follows a weighted distribution, $\widetilde{N} \sim \frac{n p_{k(n)}}{\bar{m}_k}$ ($n\in\nbbZ^+$) \cite{chiu2013stochastic}.  Now, starting from \eqref{eq::sum::product::defn},
\begin{align*}
&\int\limits_{\ncalN}\sum\limits_{{\bf x\in\psi}}g({\bf x})\prod\limits_{{\bf y}\in\psi\setminus\{{\bf x}\}}v({\bf x},{\bf y})P({\rm d }\psi)\\
&\myeq{a}{\sum\limits_{n=1}^{\infty}\int\limits_{{\ncalN}_n}\sum\limits_{{\bf x}\in\psi}g({\bf x})\prod\limits_{\substack{{\bf y}\in\psi\setminus\{{\bf x}\}}} v({\bf x},{\bf y})P({\rm d}{\psi})}
\\&=\sum\limits_{n=1}^{\infty}\ \idotsint\limits_{[{\bf x}_1,\dots,{{\bf x}_n}]\in\nbbR^{2n}}\sum\limits_{i=1}^ng({\bf x}_i)\bigg[\prod\limits_{\substack{j=1,\\j\neq i}}^{n}v({\bf x}_i,{\bf x}_j)f_k({\bf x}_j-{\bf z}){\rm d}{\bf x}_j\bigg] \\&\times f_k({\bf x}_i-{\bf z}){\rm d}{\bf x}_i \frac{np_{k(n)}}{\bar{m}_k}
\\&=\sum\limits_{n=1}^{\infty}n\int\limits_{\nbbR^2}g({\bf x}) \left(\ \int\limits_{\nbbR^2}v({\bf x},{\bf y})f_k({\bf y}-{\bf z}){\rm d}{\bf y}\right)^{n-1}\\&\qquad\times f_k({\bf x}-{\bf z}){\rm d}{\bf x}\: n\frac{p_{k(n)}}{\bar{m}},
\end{align*}
where  $\ncalN$ denotes the space of locally finite and simple point sequences in $\nbbR^2$. In (a), $\ncalN$ is partitioned into $\{\ncalN_{n}: n\geq 1 \}$ where $\ncalN_n$ is the collection of point sequences having $n$ points. This completes the proof. 
\end{IEEEproof}
\subsection{PGFL of a Cluster of PCP}
We assume that $\Psi$ is a cluster of $\Phi_k$ ($k\in\ncalK_2$) centered at ${\bf z}$.
 We present the expressions of the PGFLs of $\Psi$ with respect to its original and reduced Palm distribution in the following Lemmas.   
\begin{lemma}\label{lemm::pgfl::cluster}
 The PGFL  of $\Psi$ can be expressed as: 
\begin{multline}\label{eq::pgfl::reduced::cluster}
{G}_c(v({\bf x},{\bf y})|{\bf z})=\nbbE\left[\prod\limits_{{\bf y}\in\Psi}v({\bf x},{\bf y})\right]\\=\ncalM\bigg(\ \int\limits_{\nbbR^2}v({\bf x},{\bf y}) f_k({\bf y}-{\bf z}){\rm d}{\bf y}\bigg),
\end{multline}
where $\ncalM(z)$ is the PGFL of the number of points in $\Psi$, i.e., $N_k$.
\end{lemma}
\begin{IEEEproof}
The proof directly follows the definition of PGFL and is skipped. 
\end{IEEEproof}
\begin{lemma}\label{lemm::pgfl::reduced::cluster}
 The PGFL  of $\Psi$  under its reduced Palm distribution is given by:
\begin{multline}\label{eq::pgfl::cluster}
\widetilde{G}_c(v({\bf x},{\bf y})|{\bf z})=
\nbbE\left[\prod\limits_{{\bf y}\in\Psi\setminus\{{\bf x}\}}v({\bf x},{\bf y})\right]\\=\sum\limits_{n=1}^{\infty}\bigg(\ \int\limits_{\nbbR^2}v({\bf x},{\bf y}) f_k({\bf y}-{\bf z}){\rm d}{\bf y}\bigg)^{n-1}\:n\frac{p_{n(k)}}{\bar{m}_k}.
\end{multline}
 
\end{lemma}
\begin{IEEEproof}
The PGFL of $\Psi$ can be written as:
\begin{align*}
&\widetilde{G}_c(v({\bf x},{\bf y})|{\bf z})  = \int\limits_{\ncalN}\prod\limits_{{\bf y}\in\psi}v({\bf x},{\bf y})P^!_{\bf x}({\rm d }\psi)\\
&  \myeq{a}\sum\limits_{n=1}^{\infty}\int\limits_{{\ncalN}_n}\prod\limits_{{\bf y}\in\psi\setminus\{{\bf x}\}}^n v({\bf x},{\bf y})P({\rm d}{\psi})
\\&=\sum\limits_{n=1}^{\infty}\left(\ \int\limits_{\nbbR^2}v({\bf x},{\bf y})f_k({\bf y}-{\bf z}){\rm d}{\bf y}\right)^{n-1}\: n\frac{p_{n(k)}}{\bar{m}_k}.
\end{align*}
Note that we have partitioned $\ncalN$ in the same way as we did in the proof of Lemma~\ref{lemm::sumproduct::finite}. Since we condition on a point ${\bf x}$ of $\Psi$ to be removed, it implies that $\Psi$ will have at least one point. Hence, the number of points in $\Psi$ will follow the weighted distribution: ${\tilde{N}}\sim \frac{np_{n(k)}}{\bar{m}_{k}}$ (as was the case in Lemma~\ref{lemm::sumproduct::finite}).
\end{IEEEproof}

\subsection{Coverage Probability}We now provide our main result for coverage probability in the following Theorem. 
\begin{theorem}\label{thm::coverage} Assuming that the typical user connects to the BS providing maximum $\sir$ and $\beta_k>1,\ \forall\ k\in\ncalK$, coverage probability can be expressed as follows:

\vspace{-0.1in}
{\small
\begin{align}\label{eq::coverage_main_result}
&\pc =\nbbE\bigg[\sum\limits_{{\bf x}\in\Phi_0}\prod\limits_{j\in\ncalK\setminus\{0\}}G_j(v_{0,j}({\bf x},{\bf y}))\prod\limits_{{\bf y}\in \Phi_0\setminus
\{{\bf x}\}}v_{0,0}({\bf x},{\bf y})\bigg]+\notag
\\&\sum\limits_{k\in\ncalK_1}\int\limits_{\nbbR^2} \widetilde{G}_k(v_{k,k}({\bf x},{\bf y}))\prod_{j\in\ncalK\setminus\{k\}}{G}_j(v_{k,j}({\bf x},{\bf y}))\Lambda_k({\rm d}{\bf x})+ \notag
\\&\sum\limits_{k\in\ncalK_2}\ \iint\limits_{\nbbR^2\times\nbbR^2} \widetilde{G}_k(v_{k,k}({\bf x},{\bf y})|{\bf z})\prod_{j\in\ncalK\setminus\{k\}}{G}_j(v_{k,j}({\bf x},{\bf y}))\Lambda_k({\rm d}{\bf x}, {\rm d} {\bf z}), 
\end{align}
}
where
\begin{align}\label{eq::v_function}
v_{i,j}({\bf x},{\bf y}) = \frac{1}{1+\beta_i\frac{P_j}{P_i}\big(\frac{\|{\bf x}\|}{\|{\bf y}\|}\big)^{\alpha}},
\end{align}
  $\Lambda_k( {\rm d}{\bf x})=\lambda_k {\rm d} {\bf x} \,(k\in {\cal K}_1) $, and $\Lambda_k( {\rm d}{\bf x}, {\rm d} {\bf z})=\lambda_{{\rm p}_{k}}\bar{m}_kf_k({\bf x}-{\bf z})\:{\rm d}{\bf z}\:{\rm d}{\bf x}\, (k\in {\cal K}_2)$. Here, $G_k(\cdot)$ and $\widetilde{G}_k({\cdot})$ denote the PGFLs of $\Phi_k$ with respect to its original and reduced Palm distribution.
\end{theorem}
\begin{IEEEproof}
See Appendix~\ref{app::thm::coverage}.
\end{IEEEproof}
We observe that $\pc$ is the summation of $(K+1)$ terms due to the contribution of $(K+1)$ tiers in $\Phi$. Except the first term, the  rest $K$ terms ($k\in\ncalK\setminus\{0\}$) are in form of the product of PGFLs of $\Phi_k$ ($k\in\ncalK$) integrated with respect to the intensity measure due to the application of Lemmas \ref{lemm::sumproduct::ppp} and \ref{lemm::sumproduct::pcp}. The first term will be handled separately in Lemma~\ref{lemm::tier0} for different cases corresponding to the user distributions (cases 1-3 defined in Section~\ref{sec::system_model}). 
  In the following Lemmas, we evaluate the PGFLs of $\Phi_j$ ($j\in \ncalK$) with respect to the original and reduced palm distributions  
of $\Phi_k$ ($k\in\ncalK$), at $v_{k,j}({\bf x},{\bf y})$ given by \eqref{eq::v_function}, which can be directly substituted in \eqref{eq::coverage_main_result} to obtain the final expression of $\pc$. While Theorem \ref{thm::coverage} is general and applicable for any PCP, we specialize our results for Neyman-Scott processes hereafter.%

\begin{lemma}\label{Lem: Laplace tier j not equal to zero}   The PGFL of $\Phi_j$ ($j\in \ncalK\setminus\{0\}$) evaluated at $v_{k,j}({\bf x},{\bf y})$  is expressed as:
\begin{multline}
{G}_j(v_{k,j}({\bf x},{\bf y}))=\exp\bigg(-\pi \lambda_j \left(\frac{P_j \beta_k}{P_k}\right)^{\frac{2}{\alpha}} \|{\bf x}\|^2  C(\alpha)\bigg); \,\\ \forall j\in {\cal K}_1,
\end{multline}

\begin{multline}
{G}_j(v_{k,j}({\bf x},{\bf y}))=  \exp \Big(-\lambda_{{\rm p}_k} \int_{\nbbR^2} \big(1-   \mu_{k,j}({\bf x }, {\bf z})  \big) {\rm d} {\bf z}  \Big); \\ \forall j\in {\cal K}_2,
\end{multline}
with $ \mu_{k,j}({\bf x }, {\bf z}) = \exp \Big( -\bar{m}_j  \int_{\nbbR^2}     \Big(\frac{1}{1+\frac{P_k \|{\bf y} \|^{\alpha}}{ \beta_k P_j \|{\bf x}\|^{\alpha}}}  \Big)   f_j({\bf y}-{\bf z}) {\rm d} {\bf y} \Big)$ and $C(\alpha)=\frac{1}{\sinc(\frac{2}{\alpha})}$.
\end{lemma}
\begin{IEEEproof}
See Appendix~\ref{app: Laplace tier j not equal to zero}. 
\end{IEEEproof}
In the next Lemma, we characterize the PGFL of $\Phi_0$. Note that this case is handled separately since the definition of  $\Phi_0$ depends on the user configurations to be considered. 
\begin{lemma}\label{lemm::pgfl_phi_0} The PGFL of  $\Phi_0$ is given by:
\begin{itemize}
\item case 1:\ ${G}_0(v_{k,0}({\bf x},{\bf y}))=1,$
\item case 2:\ $ {G}_0(v_{k,0}({\bf x},{\bf y}))=  \int_{\nbbR^2}\frac{1}{1+   \frac{ P_0 \beta_k}{P_k} \|{\bf x}\|^{\alpha} \|{\bf y}\|^{-\alpha} }  f_0({\bf y}) {\rm d}{\bf  y},$
\item case 3: ${G}_0(v_{k,0}({\bf x},{\bf y})|{\bf z})= \mu_{k,0}({\bf x }, {\bf z}),$
\end{itemize}
where $\mu_{k,0}({\bf x }, {\bf z})$ is given by Lemma~\ref{Lem: Laplace tier j not equal to zero}.
\end{lemma}
\begin{IEEEproof}
See Appendix~\ref{app::pgfl_phi_0}.
\end{IEEEproof}
Having characterized the expressions of PGFLs evaluated at $v_{k,j}({\bf x},{\bf y})$ $\forall j\in \ncalK$, we now focus on the evaluation of the  PGFL of $\Phi_k$ with respect to its reduced Palm distribution.
\begin{lemma} \label{lemm::pgfl_reduced}
The  PGFL of $\Phi_k$ ($k\in\ncalK\setminus\{0\}$) under its reduced Palm distribution is given by:
\begin{align}
\widetilde{G}_k(v_{k,k}({\bf x},{\bf y}))={G}_k(v_{k,k}({\bf x},{\bf y})); \qquad \text{when $k\in\ncalK_1 $},
\end{align}
\begin{multline}
\widetilde{G}_k(v_{k,k}({\bf x},{\bf y})|{\bf z})= \exp \Big(-\lambda_{{\rm p}_k} \int_{\nbbR^2} \big(1-     \mu_{k,k}({\bf x }, {\bf z}')   \big) {\rm d} {\bf z}'  \Big) 
 \\  \exp\Big( -\bar{m}_k     \Big(  \int_{\nbbR^2}  (1-v_{k,k}({\bf x},{\bf y}))    f_k({\bf y}-{\bf z})  {\rm d} {\bf y}\Big) \Big) 
 ; \:    \text{when $ k\in {\cal K}_2$},
\end{multline}
where ${G}_k(v_{k,k}(\cdot)) $ and  $\mu_{k,k}(\cdot)$  are given by Lemma~\ref{Lem: Laplace tier j not equal to zero}.
\end{lemma}
\begin{IEEEproof}See Appendix~\ref{app::pgfl_reduced}. 
\end{IEEEproof}
Referring to \eqref{eq::coverage_main_result}, we denote the first term under summation as:
\begin{align}
\pc_0 = \nbbE\bigg[\sum\limits_{{\bf x}\in\Phi_0}\prod\limits_{j\in\ncalK\setminus\{0\}}G_j(v_{0,j}({\bf x},{\bf y}))\prod\limits_{{\bf y}\in \Phi_0\setminus
\{{\bf x}\}}v_{0,0}({\bf x},{\bf y})\bigg]. 
\end{align}
This term is now evaluated for the three cases next.
\begin{lemma}\label{lemm::tier0}
 If $\Phi_0=\emptyset$ (case 1) then,
\begin{align}
{\tt P}_{{\rm c}_0}=0.
\end{align}
If $\Phi_0=\{{\bf z}\}$ (case 2) then,
\begin{align}
{\tt P}_{{\rm c}_0}=\int_{\nbbR^2} \prod\limits_{j\in\ncalK\setminus\{0\}}G_j(v_{0,j}({\bf x},{\bf y})) f_0({\bf x}) {\rm d} {\bf x}.
\end{align}
If $\Phi_0=\ncalB_i^{{\bf z}_0}$ (case 3)  then ${\tt P}_{{\rm c}_0} = $
\begin{align}\notag
&\int_{\nbbR^2}  \int_{\nbbR^2} \exp \bigg(-\bar{m}_0 \bigg( \int_{\nbbR^2}    \Big( 1-   v_{0,0}({\bf x},{\bf y}) \Big)  f_0({\bf y}-{\bf z})  {\rm d} {\bf y} \bigg)\bigg) \notag
\\&\times \Big(\bar{m}_0 \int_{\nbbR^2} v_{0,0}({\bf x},{\bf y})  f_0({\bf y}-{\bf z})  {\rm d} {\bf y}   +1 \Big) \notag
\\& \times \prod\limits_{j\in\ncalK\setminus\{0\}}G_j(v_{0,j}({\bf x},{\bf y})) f_0({\bf x}-{\bf z}) f_0({\bf z}) \:{\rm d} {\bf x}\: {\rm d} {\bf z},
\end{align}
where $G_j(\cdot)$ is given by Lemma~\ref{lemm::pgfl_phi_0}.
\end{lemma}
\begin{IEEEproof}
See Appendix~\ref{app::tier0}.
\end{IEEEproof}

\begin{figure}
  \centering{
  \includegraphics[width=.37\textwidth]{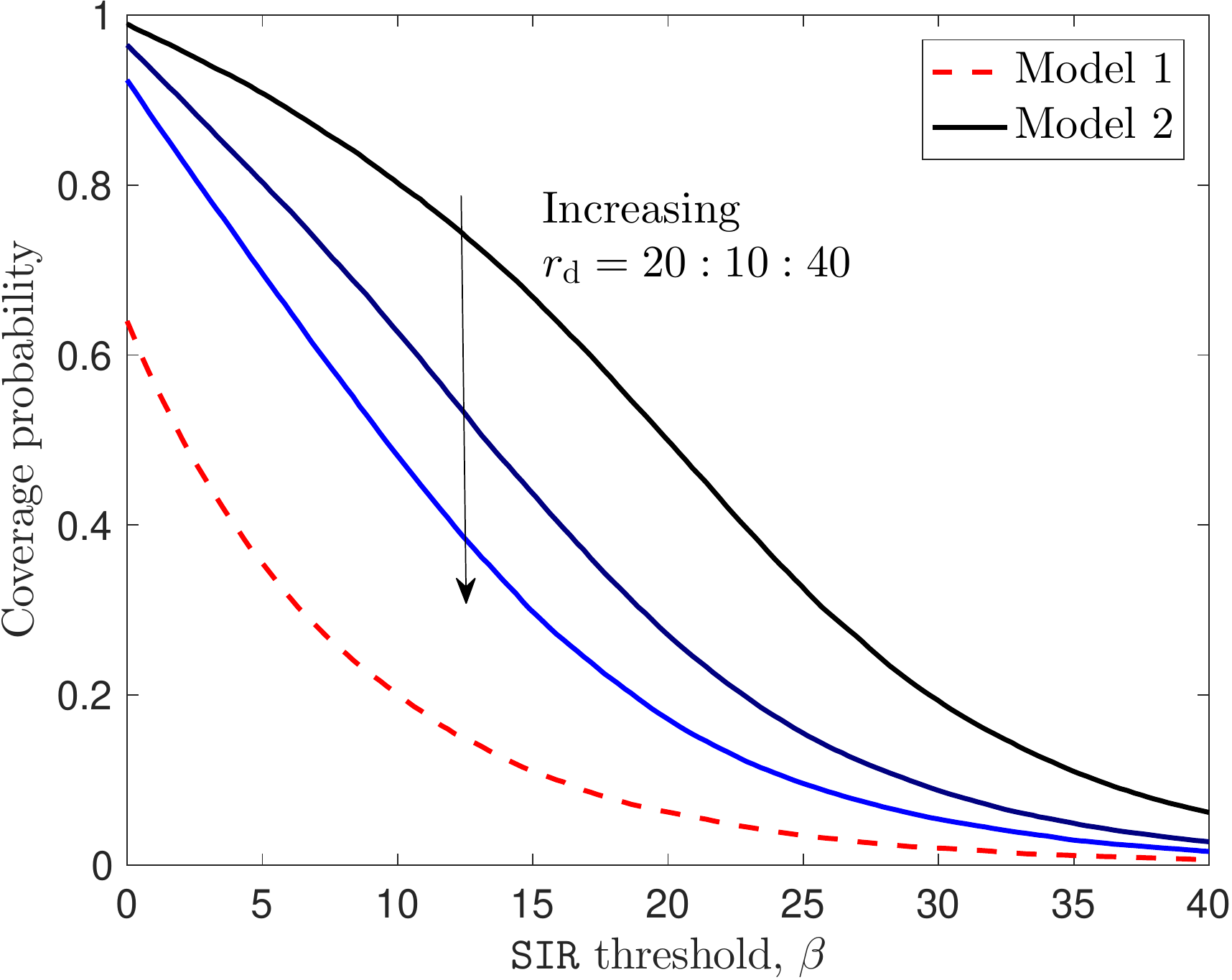}
          
                }
                \caption{ \small Comparison of the coverage probability in  Model 1 and 2 ($\alpha=4$, $P_{\rm m}=1000 P_{\rm s}$, $\lambda_{\rm s}= 100 \lambda_{\rm s}$, and  $\beta_{\rm s}= \beta_{\rm m}= \beta$).}
                                \label{Fig:  Model2}

\end{figure}

\begin{figure}
  \centering{
  \includegraphics[width=.37\textwidth]{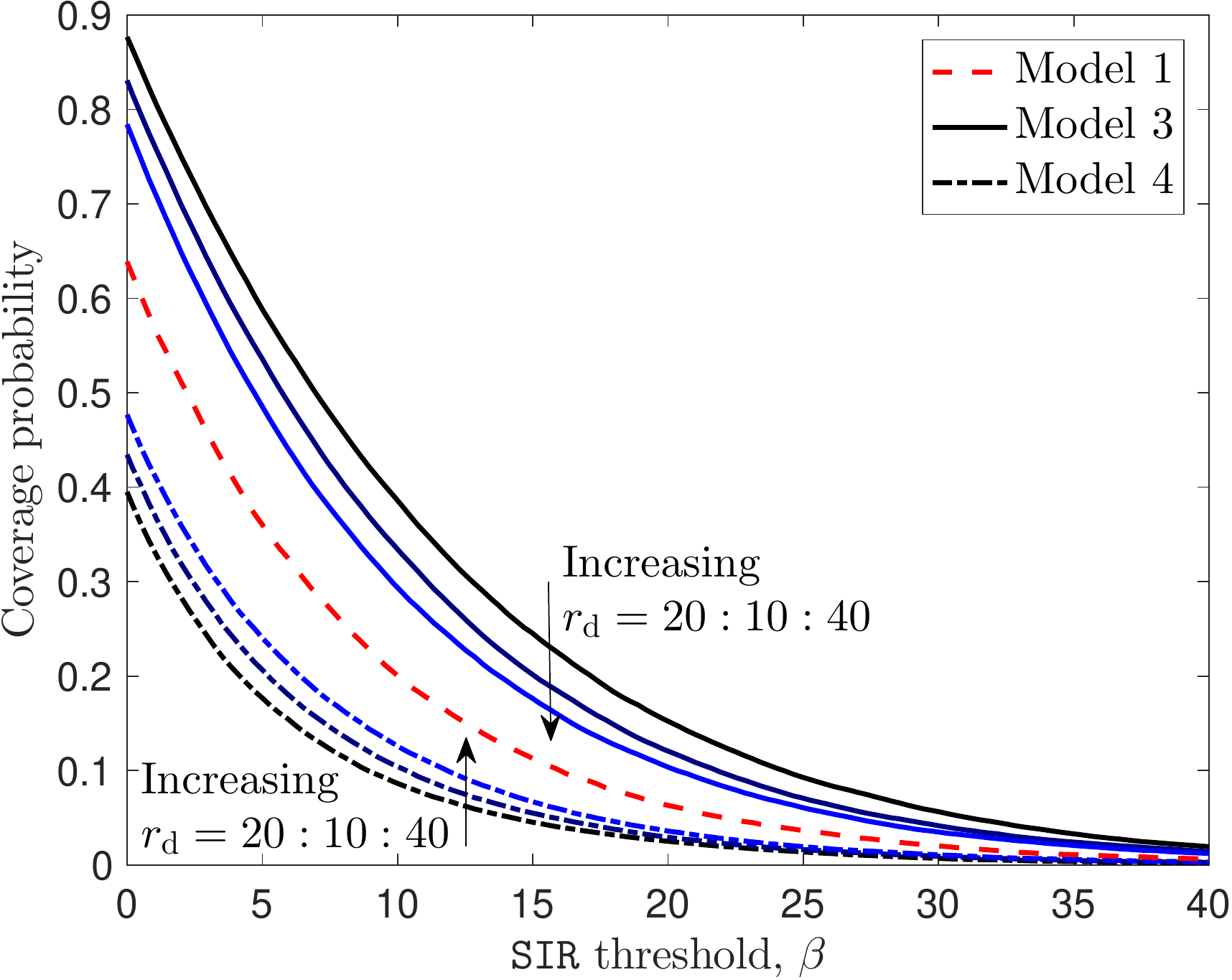}
          
                }
                \caption{\small Comparison of the coverage probability in  Model 1,  3, and  4  ($\alpha=4$, $P_{\rm m}=1000 P_{\rm s}$,  $\bar{m}=3$, and $\beta_{\rm s}= \beta_{\rm m}= \beta$).}
                                \label{Fig:  Model3and4}

\end{figure}

\section{Results and Discussion}
In this section, we compare the performance of  Models 1-4  in terms of the coverage probability. For this comparison, we assume that MBSs  transmit at fixed power $P_{\rm m}$ and distributed as a PPP of density $\lambda_{\rm m}$ in all four models.  In \figref{Fig:  Model2}, we compare the coverage probability of   Models 1 and  2, where SBSs are 
spatially distributed as a PPP with density $\lambda_{\rm s}$ and transmit at power $P_{\rm s}$.
 Following the notion of ``clustered'' users in 3GPP models, the non-uniformly distributed users  
are assumed to be a realization of a \matern cluster process (MCP) in Model 2.
   This means that users are assumed to be uniformly distributed within a disk of radius $r_{\rm d}$ around SBSs. As evident from \figref{Fig:  Model2}, the coverage probability decreases as $r_{\rm d}$ increases and   converges towards  that of  Model 1. The limiting nature of the coverage probability  and its convergence to Model 1 as cluster radius goes to infinity is formally proved in~\cite{SahaAfshDh2016}, where the typical user is served by the BS that provides maximum received power averaged over fading. The reason of the coverage boost for denser cluster is that the SBS at cluster center lies closer to the typical user with high probability, hence improving the signal quality of the serving link.

Next, we plot coverage probability of  Models 1,  3, and 4 in \figref{Fig:  Model3and4}. 
In Model 3, user and SBS locations are two independent realizations of an MCP conditioned on its parent PPP. More precisely, user and SBS clusters are colocated around the same set of cluster centers and have  the same cluster radius, i.e., $r_{\rm d}$.  While SBSs in Model 4 are also assumed to be  realizations of an MCP,  users are independently  distributed in $\nbbR^2$. From \figref{Fig:  Model3and4}, it can be deduced that increasing $r_{\rm d}$ has a conflicting effect on coverage probability of Model 3 and 4: coverage probability of Model 4 increases whereas that of  Model 3 decreases. For Model 3, as $r_{\rm d}$ increases,  the collocated user and SBS clusters become sparser and the candidate serving SBS lies farther to the typical user with high probability. On the contrary, for Model 4 where the users locations are independent and uniform over space,  the distance between the candidate serving SBS and the typical user decreases more likely with the increment of $r_{\rm d}$.

\section{Conclusion}
In this paper, we developed a unified HetNet model by combining PPP and PCPs that accurately models variety of spatial configurations for SBSs and users considered in the 3GPP simulation models. This is a significant generalization of the PPP-based $K$-tier HetNet model of \cite{5743604,dhillon2012modeling}, which was not rich enough to model non-uniformity and coupling across the locations of users and SBSs. For this model, we characterized the downlink coverage probability under max-$\sir$ cell association. As a part of our analysis, we evaluated the sum-product functional for PCP and the associated offspring point process.
This work has numerous extensions. An immediate extension is the coverage probability analysis with the relaxation of the assumption on $\sir$-thresholds ($\beta_k$) being greater than unity. From stochastic geometry perspective, this will necessitate the characterization of the $n$-fold Palm distribution of PCP and its offspring point process. Extensions from the cellular network perspective involve analyzing other metrics like rate and spectral efficiency in order to obtain further insights into the network behavior. Coverage probability analysis under this setup for uplink is another promising future work. From modeling perspective, we can incorporate more realistic channel models e.g. shadowing and general fading.

\appendix
\subsection{Proof of Theorem~\ref{thm::coverage}}\label{app::thm::coverage}
Under the assumption that $\beta_k>1,\ \forall\ k\in\ncalK$, there will be at most one BS $\in\Phi$ satisfying the condition for coverage \cite{dhillon2012modeling}. 
Continuing from \eqref{eq::coverage_definition},
{\small
\begin{align}
&\pc=\sum\limits_{k\in\ncalK}\nbbE\bigg[\sum\limits_{{\bf x}\in\Phi_k}  {\bf 1}\bigg( \frac{P_k h_{\bf x}\|{\bf x}\|^{-\alpha}}{\ncalI(\Phi_k\setminus\{{\bf x}\})+\sum\limits_{j\in\ncalK\setminus\{k\}}\ncalI(\Phi_j)}>\beta_k\bigg)\bigg]\notag\\
&=\sum\limits_{k\in\ncalK}\nbbE\bigg[\sum\limits_{{\bf x}\in\Phi_k}\nbbP\big(h_{\bf x}>\frac{\beta_k}{{P_k}}\big(\ncalI(\Phi_k\setminus\{{\bf x}\})+\sum\limits_{j\in\ncalK\setminus\{k\}}\ncalI(\Phi_j)\big)\|{\bf x}\|^{\alpha}\big)\bigg]\notag\\
&\myeq{a}\sum\limits_{k\in\ncalK}\nbbE\bigg[\sum\limits_{{\bf x}\in\Phi_k}\exp\big(-\frac{\beta_k}{P_k}\big(\ncalI(\Phi_k\setminus\{{\bf x}\})+\sum\limits_{j\in\ncalK\setminus\{k\}}\ncalI(\Phi_j)\big)\|{\bf x}\|^{\alpha}\big)\bigg]\notag\\
&= \sum\limits_{k\in\ncalK}\nbbE\bigg[\sum\limits_{{\bf x}\in\Phi_k}\exp\bigg(-\frac{\beta_k}{P_k}(\ncalI(\Phi_k\setminus\{{\bf x}\})\bigg)\Theta_k({\bf x})\bigg].\label{eq::coverage::intermediate}
\end{align}
}
Here, step (a) follows from $h_{\bf x}\sim \exp(1)$. The final step   follows from the independence  of $\Phi_k$, $\forall\ k\in\ncalK$, where,
\begin{align*}
&\Theta_k({\bf x})=\prod\limits_{j\in\ncalK\setminus\{k\}}\nbbE\exp\left(-\frac{\beta_k}{P_k}
\ncalI(\Phi_j)\|{\bf x}\|^{\alpha}\right)\\
&=\prod\limits_{j\in\ncalK\setminus\{k\}}\nbbE\exp\left(-{\frac{\beta_k\|{\bf x}\|^{\alpha}}{P_k}\sum\limits_{{\bf y}\in\Phi_j} P_j{h_{\bf y}}\|{\bf y}\|^{-\alpha}}\right)\\
&=
\prod\limits_{j\in\ncalK\setminus\{k\}}\nbbE\prod\limits_{{\bf y}\in\Phi_j}\nbbE_{h_{\bf y}}\exp\left(-{\frac{\beta_k\|{\bf x}\|^{\alpha}}{P_k}{P_j h_{\bf y}}\|{\bf y}\|^{-\alpha}}\right)\\
&\myeq{a}  \prod\limits_{j\in\ncalK\setminus\{k\}}\nbbE\prod\limits_{{\bf y}\in\Phi_j}\frac{1}{1+ \beta_k \frac{ P_j}{P_k}\left(\frac{\|{\bf x}\|}{\|{\bf y}\|}\right)^{\alpha}}\\
&=\prod\limits_{j\in\ncalK\setminus\{k\}}G_j(v_{k,j}({\bf x},{\bf y})).
\end{align*}
Step (a) follows from the fact that $\{h_{\bf y}\}$ is an i.i.d. sequence of exponential random variables. 
Following from \eqref{eq::coverage::intermediate}, we get,
\begin{align*}
\pc&=  \sum\limits_{k\in\ncalK}\nbbE\bigg[\sum\limits_{{\bf x}\in\Phi_k}\Theta_k({\bf x})\exp\bigg(-\frac{\beta_k}{P_k}\ncalI(\Phi_k\setminus\{{\bf x}\})\bigg)\bigg]\\
&=\sum\limits_{k\in\ncalK}\nbbE\bigg[\sum\limits_{{\bf x}\in\Phi_k}\Theta_k({\bf x})\prod\limits_{{\bf y}\in \Phi_k\setminus
\{{\bf x}\}}v_{k,k}({\bf x},{\bf y})\bigg].
\end{align*}
The exponential term can be simplified following on similar lines as that of $\Theta_k({\bf x})$. 

Thus $\pc$ can be written as the summation of $K+1$ terms each in sum-product form defined in \eqref{eq::sum::product::defn}.  For $k\in\ncalK_1$ and $k\in\ncalK_2$,
the final result  is obtained  by direct application of Lemmas~\ref{lemm::sumproduct::ppp} and \ref{lemm::sumproduct::pcp}, respectively.
\subsection{Proof of Lemma~\ref{Lem: Laplace tier j not equal to zero}}\label{app: Laplace tier j not equal to zero}
When $j\in\ncalK_1$, $G_j(v_{k,j}({\bf x},{\bf y}))$ is the PGFL of PPP which is given by \cite[Theorem 4.9]{haenggi2012stochastic}:
\begin{align}\label{eq::PGFL_PPP}
G_j(v_{k,j}({\bf x},{\bf y}) ) = \exp\left(-\int\limits_{\nbbR^2}(1-v_{k,j}({\bf x},{\bf y}))\lambda_j{\rm d}{\bf y}\right).
\end{align}
When $k\in\ncalK_2$, $G_j(v_{k,j}({\bf x},{\bf y}))$ is the PGFL of PCP which is given by \cite[Theorem 4.9]{haenggi2012stochastic}:
\begin{multline}\label{eq::PGFL_PCP}
G_j(v_{k,j}({\bf x},{\bf y}) ) = \exp\bigg(-\lambda_{{\rm p}_k}\int\limits_{\nbbR^2}\bigg(1-{\cal M}\bigg( \int_{\nbbR^2}v_{k,j}({\bf x},{\bf y})\\\times
f_{j}({\bf y}-{\bf z}){\rm d}{\bf y}\bigg){\rm d}{\bf z} \bigg)\bigg),
\end{multline}
where $\ncalM(z)=\nbbE(z^{N_j}) = \exp(-\bar{m}_j(1-z))$ is the moment generating function of $N_j$ ($j\in\ncalK_2$). Finally we substitute $v_{k,j}({\bf x},{\bf y})$ given by \eqref{eq::v_function} to obtain the desired expressions. 
\subsection{Proof of Lemma~\ref{lemm::pgfl_phi_0}}\label{app::pgfl_phi_0}
In case 1,  $\Phi_0$ is a null set if users are distributed according to a PPP, and hence $G_0(v_{k,0}({\bf x},{\bf y}))=1$. In case 2, where users are distributed as a PCP with parent PPP $\Phi_j$ $(j\in \ncalK_1)$,  
\begin{align}
G_0(v_{k,0}({\bf x},{\bf y}))=\int\limits_{\nbbR^2}v_{k,0}({\bf x},{\bf y})f_0({\bf y}){\rm d}{\bf y}.
\end{align}
 In case 3,   $\Phi_{0}$ is a  cluster of $\Phi_j$ ($j\in\ncalK_2$) centered at $\bf z$. Its  PGFL is provided by Lemma~\ref{lemm::pgfl::cluster} with the substitution $\ncalM(z)= \exp(-\bar{m}_j(1-z))$ for Neyman Scott process. 
\subsection{Proof of Lemma~\ref{lemm::pgfl_reduced}}
\label{app::pgfl_reduced}
When $k\in\ncalK_1$, $\Phi_k$ is a PPP and its reduced palm distribution is the  same as its original distribution (Slivnyak's theorem, \cite{chiu2013stochastic}). However, this is not true for PCP (when $k\in\ncalK_2$). Denote by 
${\cal A}_k={\cal B}^{\bf z}_{k}+{\bf z}$,  the cluster within  which the serving BS is located.  The PGFL with respect to the reduced palm distribution of a PCP can be derived as: 
\begin{align*}
&\widetilde{G}_k(v_{k,k}({\bf x},{\bf y})|  {\bf z}) =\E \Big[ \prod_{{\bf y} \in \Phi_k  \setminus \{ {\bf  x}\}}   v_{k,k}({\bf x},{\bf y})    \Big]\myeq{a} {G}_k(v_{k,k}({\bf x},{\bf y})) \\&\times \E \big[\prod_{{\bf y} \in {\cal A}_k \setminus \{\bf x\}} v_{k,k}({\bf x},{\bf y})\big] 
={G}_k(v_{k,k}({\bf x},{\bf y}))  \widetilde{G}_{c}(v_{k,k}({\bf x},{\bf y})|{\bf z}), 
\end{align*}
where (a) follows from Slivnyak's theorem and definition of PGFL. The final result follows from Lemma~\ref{lemm::pgfl::reduced::cluster} along with the fact that $N_k = \mathtt{Poisson}({\bar{m}_k})$  for Neyman Scott  process. 
\subsection{Proof of Lemma~\ref{lemm::tier0}}
\label{app::tier0}
Case 1 is trivial. For case 2, $\Phi_0$ has only one point with distribution $f_0({\bf z})$. For case 3, we use Lemma~\ref{lemm::sumproduct::finite} where $g({\bf x })=\prod\limits_{j\in\ncalK\setminus\{0\}}G_j(v_{0,j}({\bf x},{\bf y}))$ and $v({\bf x},{\bf y})=v_{0,0}({\bf x},{\bf y})$ 
and substitute $p_{k(n)}$ follows Poisson distribution. Hence,
\begin{align*}
{\tt P}_{{\rm c}_0}&=  \int_{\nbbR^2}  \int_{\nbbR^2}\sum_{n=1}^{\infty} n\bigg( \int_{\nbbR^2}     v_{0,0}({\bf x},{\bf y})   f_0({\bf y}-{\bf z})  {\rm d} {\bf y} \bigg)^{n-1} 
\\& \prod\limits_{j\in\ncalK\setminus\{0\}}G_j(v_{0,j}({\bf x},{\bf y}))\frac{n e^{-\bar{m}_0}}{\bar{m}_0 n!}  f_0({\bf x}-{\bf z}) f_0({\bf z})\:{\rm d} {\bf x}\: {\rm d} {\bf z}. 
\end{align*}
\balance
\bibliographystyle{IEEEtran}
\bibliography{ITA_prototype_2_v8-3.bbl}

\end{document}